\documentstyle[12pt,twoside,cosmion,epsf,psfig,rotate]{article}
\heads{High Energy Astrophysics: Accretion Theory, Gamma Ray Bursts,  
Relativistic and Compact Objects} 
{Stochastic spin evolution of neutron stars \ \ 
{\rm S.B. Popov et al.}}
\begin{document}

\Arthead{1}{1}

\Title{Stochastic spin evolution of neutron stars}
{S.B. Popov$^1$, M.E. Prokhorov$^1$, A.V.
Khoperskov$^2$, V.M. Lipunov$^{1,3}$}
{$^1$Sternberg Astronomical Institute,
              Universitetski pr. 13, 119899 Moscow\\
$^2$Volgograd State University, Department of Theoretical Physics, 
40068 Volgograd\\
$^3$Moscow State University, Department of Physics}

\Abstract{
 In this paper we present calculations of period distribution for
old accreting isolated neutron stars (INSs).
 
At the age about few billions years 
low velocity INSs come to the stage of accretion. At that stage their
period evolution is governed by magnetic braking and
accreted angular momentum.
Due to turbulence of the interstellar medium (ISM) accreted momentum
can both accelerate and decelerate rotation of an INS and 
spin evolution has chaotic character.

 Calculations show that for constant magnetic field INSs
have relatively long spin periods, $\ge 10^4$--$10^5$ s,
depending on parameters of INSs and ISM density.
Due to long periods INSs have high spin up/spin down rates,
which should fluctuate on a time scale about few years.
}

\section{Introduction}

 Spin period is the most precisely determined
parameter of a neutron star (NS). Estimates of other parameters:
masses (for isolated objects),
radii, temperatures, magnetic fields etc. are always model dependent. 
Because of that it is very important to have a clear
picture of period evolution as far as this parameter
is usually used to determine other characteristics of NSs.
Here we try to obtain distribution of 
spin periods for old accreting INSs (AINSs).

 AINSs are now a subject of interest in astrophysics
(see Treves et al.\cite{9}).
Probably few candidates are observed by ROSAT (Motch\cite{7}).

 In the next section we describe the model we use to obtain period
distributions and show results for the easiest case of ``spin equilibrium''. 
Then in the section 3 we present our main results for the
``non-equilibrium'' case and briefly discuss them in the section 4. 
Details of calculations can be found in Prokhorov et al.\cite{8}.

\section{``Spin equilibrium''}

Previous attempts to calculate typical periods of AINSs
were made by Lipunov \& Popov\cite{5} and Konenkov \& Popov\cite{3}. 
In these papers the authors do not try to obtain distributions: 
only characteristic periods of AINSs are derived. The
authors {\it assume} that AINSs are in ``spin equilibrium'', i.e.
all AINSs in these estimates
have enough time to reach the stage at which magnetic braking is
compensated by accretion of angular momentum.
We use terms ``spin equilibrium'' and ``non-equilibrium''
in quotation-marks as far as there is no real equilibrium: 
period can significantly fluctuate.
But the situation in general is similar to real period equilibrium in
close binaries (see Ghosh \& Lamb\cite{2}, Lipunov\cite{4}).

 We start with the following equations:

\begin{equation}
\frac{d\omega}{dt}= F+\Phi,\, F=-
\frac{k_t\mu^2}{IR_{co}^3}, \, \Phi\sim\frac{\dot M J}{I}.
\end{equation}
Here $I$ -- moment of inertia of a NS,
$\omega=2\pi/p$ -- spin frequency,
$\mu=B\, R_*^3$ -- magnetic moment of a NS,
$R_{co}=\left(GM/\omega^2\right)^{1/3}$ -- corotation radius,
$k_t$ -- constant of order of unity,
and $\Phi$ -- turbulent torque, $<\Phi>=0$.
$J$ is determined as $J=$Min $(v_tR_G, v_A R_A)$, where
$v_t=10^6 {\rm cm\, s}^{-1}
\left( R_G/R_t \right)^{1/3}$ -- turbulent velocity at $R=R_G$,
$R_t=2\cdot 10^{20}$ cm, $R_G=2GM/v^2$, $v^2=v_s^2+v_{sp}^2$,
$v_s$ -- sound velocity, $v_{sp}$ -- spatial velocity,
$R_A=\left(\mu^2/2\dot M \sqrt{GM} \right)^{2/7}$ -- Alvfen radius,
and $v_A=\left( GM/R_A \right)^{1/2}$ -- Keplerian velocity at the
Alfven radius. For the most reasonable parameters $J=v_t R_G$.

 Now we have to introduce a kind
of ``diffusion coefficient'', $D$, because due to 
interstellar medium (ISM) turbulence we have
a kind of diffusion in the space of frequencies.
This coefficient can be approximatelly determined as 
$  D=(1/6) \left({\dot M J}/{I} \right)^2 {R_G}/{v}$ 
(Lipunov \& Popov\cite{5}).
Here $\dot M=\pi\,  R_G^2\, \rho \, v$ -- is an accretion rate.

 We can determine an average spin frequency in the following way:
\begin{equation}
\omega_{turb}^2=
\int\displaylimits_{0}^{\infty}\omega^4e^{-V(\mid \omega \mid)/D}d\omega /
\int\displaylimits_{0}^{\infty}\omega^2e^{-V(\mid \omega \mid)/D}d\omega ,\,
V(\mid \omega \mid)=\frac{\mu^2}{3GMI}\mid \omega\mid ^3.
\end{equation}
Finally, $p_{turb}=2\pi / \omega_{turb}$.
For $J=v_t\cdot R_G$ we can write it as:
\begin{equation}
p_{turb}=3.9 \cdot 10^8 \mu_{30}^{2/3} I_{45}^{1/3} M_{1.4}^{-26/9}
n^{-2/3}v_{7}^{43/9}
R_{t_{2,20}}^{2/9}\, {\rm s}.
\end{equation}
Here $v_{7}=v/(10^7 \, {\rm cm} \, {\rm s}^{-1})$,
$R_{t_{2,20}}=R_t/(2\cdot 10^{20} {\rm cm})$.

 We note strong dependence of $p_{turb}$ and $D$ on $v$.
If information on $p$ and $\dot p$ is available the problem can be reversed,
and one can obtain an estimate of the velocity of a AINS as it was done for
stellar wind accretion in X-ray pulsars by Lipunov \& Popov\cite{6}.

\begin{figure}
\vbox{\psfig{figure=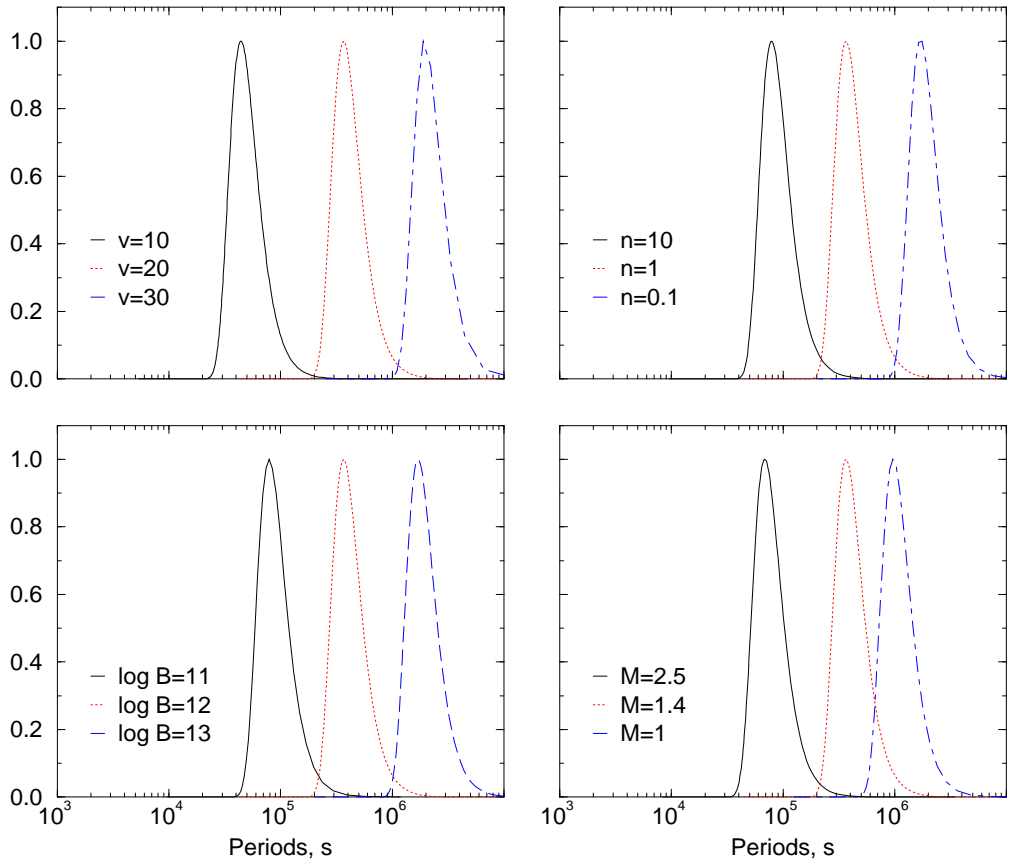,width=14.0cm,angle=-90}}
\caption[]{Period distributions in equilibrium for different 
parameters. In each of the four plots we vary one parameter:
$n, v, M, B$. Results are normalized to unity at the maximum.}
\end{figure}

Probability plotted in Fig.~1 was calculated as:

\begin{equation}
 f(\omega, v, \mu)\propto(\mu^2/GMID)\omega^2e^{-V/D},
\end{equation}
and then normalized.
Here $V$ and $D$ are functions of $v, \mu, n$.
Note, that peaks in Fig.~1 are indeed sharp. 
It means, that if ``spin equilibrium''
can be reached, it is possible to use just one typical value, $p_{turb}$,
for each set of parameters 
as it was done by Lipunov \& Popov\cite{5} and 
Konenkov \& Popov\cite{3}.

\section{``Non-equilibrium'' calculations}

In this section we calculate probability distribution for
the ``non-equilibrium'' case.
These distributions are calculated for magnetic moments
with Gaussian distribution in logarithmic scale with
central value ${\rm lg}(\mu_0)=30.06$ and $\sigma=0.32$
(see Colpi et al. 2001\cite{1} for details on magnetic evoluyion of NSs).
Velocities of AINS are taken from the Maxwellian distribution with
a mean velocity  200 km s$^{-1}$.

 We solved numerically the following differential equation:

\begin{equation}
df/dt=A/\omega^2 \partial(\omega^4\,f)/\partial\omega + 
D/\omega^2 \partial/\partial\omega
\left( \omega^2 \,\partial f/\partial\omega  \right), \, A=\mu^2/GMI.
\end{equation}

After initial parameters of an INS are chosen from
the distributions described above we check if
this NS can reach the accretion stage in $10^{10}$ yrs.
To do it we calculate time which it spends as Ejector:
$ t_E\approx 10^9 n^{-1/2} v_6 \mu_{30}^{-1} {\rm yrs},$
$v_6=v/10^6$ cm s$^{-1}$.
We neglect the stage of Propeller,
as far as for constant field it is much shorter than the stage of Ejector
(Lipunov \& Popov\cite{5}).

In Fig.~2 we present a curve for $n=1$ cm$^{-3}$ and selected INS
parameters: $\mu=2\cdot 10^{29}$ G cm$^{3}$, 
$v_{sp}=10$ km s$^{-1}$. 
In Fig.~3 we show our final results
for Maxwellian velocity distribution and log-Gaussian magnetic field
distribution for two values of the ISM density.

\begin{figure}
\hbox to\textwidth{%
\hbox to 0.45\textwidth{\hss
\epsfxsize=0.45\textwidth
{\rotate[r]{\epsfbox{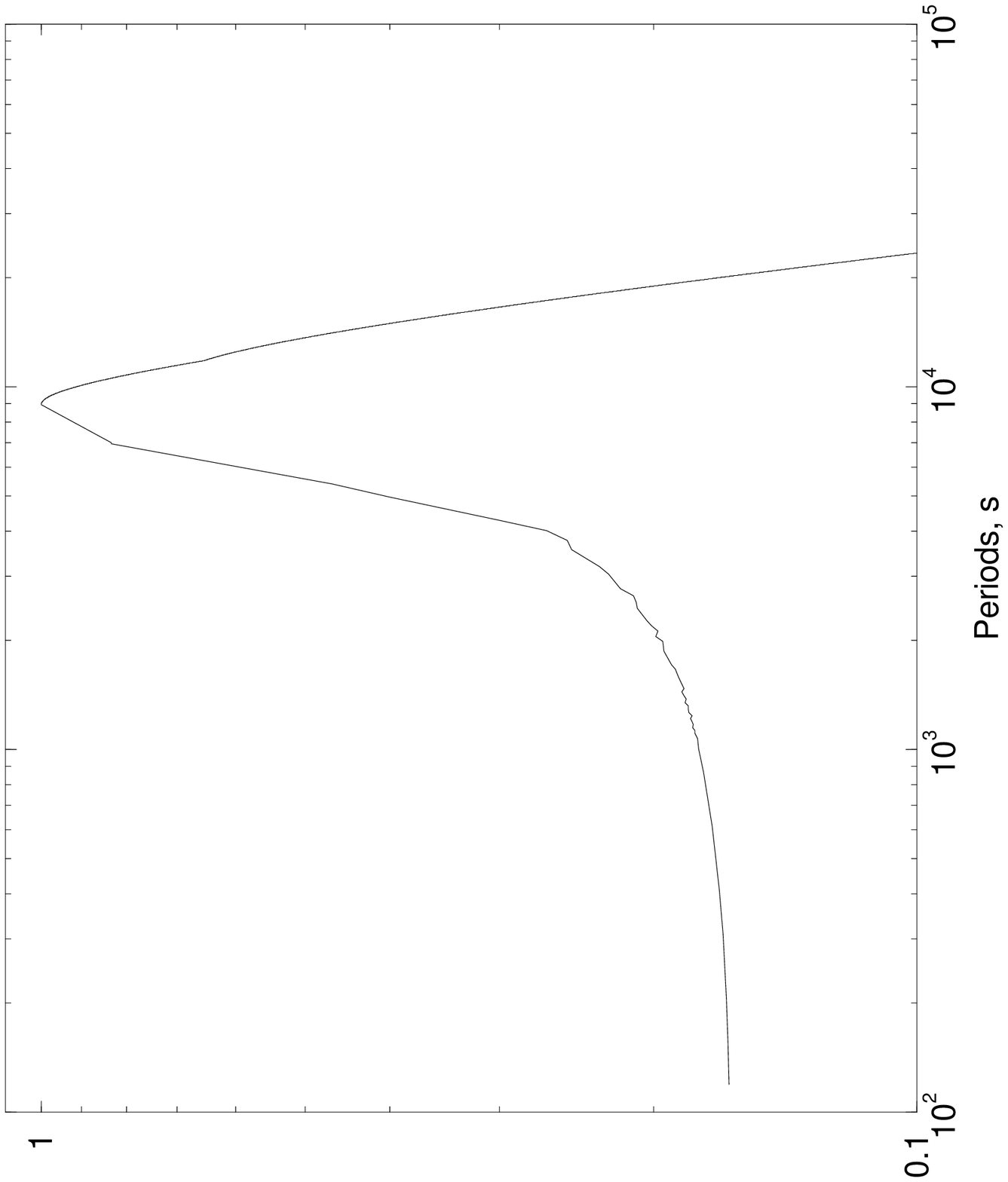}}}
\hss}
\hss
\hbox to 0.45\textwidth{\hss
\epsfxsize=0.45\textwidth
{\rotate[r]{\epsfbox{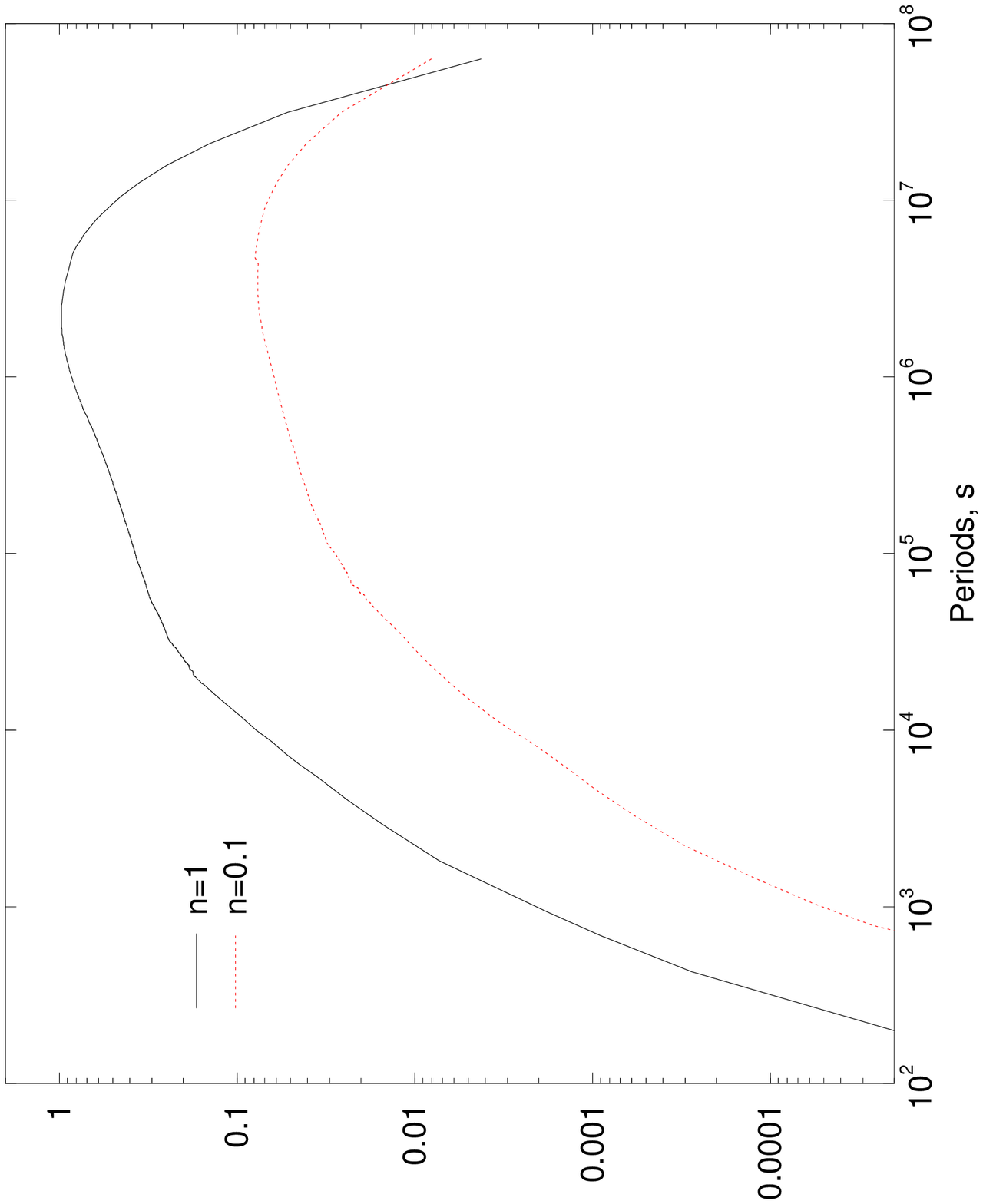}}}
\hss}
}

\hbox to\textwidth{%
\parbox{0.45\textwidth}{
\caption{Period distribution
for $\mu=2\cdot 10^{29}$ G cm$^{3}$, $v_{sp}=10$ km
s$^{-1}$, $n=1$ cm$^{-3}$. Results are normalized to unity at the maximum.}
}
\hss
\parbox{0.45\textwidth}{
\caption{Period distributions for  
$n=1$ cm$^{-3}$ and $n=0.1$ cm$^{-3}$. Results are normalized to unity
at the maximum of the highest curve.}
}
}
\end{figure}

\section{Discussion}

 After an INS come to the stage of accretion it is controlled by
two processes (see eq.1): 
magnetic spin-down and turbulent spin-up/spin-down.
Initially  magnetic spin-down is more
significant,
but at some period, $p_{cr}$, these two processes become
comparable. For longer periods an INS will be governed mainly by
turbulent forces.
One can obtain the following formula for $p_{cr}$: 
$p_{cr}^2=(4 \pi^2 \mu^2)/(GM\dot M J)$.
An INS reach $p_{cr}$ in $\Delta t\sim 10^5$--$10^7$ yrs 
after onset of accretion: 
$ \Delta t= (I \sqrt{GM})/(\mu \sqrt{\dot M J})$.

 For field decay picture should be completely different (see for example
Konenkov \& Popov\cite{3}).
AINSs with decayed field can appear as pulsating sources with
periods about 10 s and $\dot p$ about $10^{-13}$ s/s.
As an INS passes through turbulent cells 
a value and a sign of $\dot p$ will fluctuate on a time scale 
$R_G/v_{sp}\simeq 11.8\,v_{sp}/(10 \, {\rm km}\, {\rm s}^{-1})\, {\rm yr}$.\\

\noindent
{\bf Acknowledgments}

\noindent
This work was supported by RFBR (01-02-06265, 00-02-17164,
01-15(02)-99310).\\
SP and MP thank M. Colpi, A. Treves and R. Turolla for discussions.


\begin{thebibliography}{}
\bibitem{1}
Colpi M., Possenti A., Popov S.B., Pizzolato F.: 2001,
in `Physics of Neutron Star Interiors'', 
Eds. D. Blaschke, N.K. Glendenning, \& A. Sedrakian
(Springer--Verlag, Berlin), (astro-ph/0012394)
\bibitem{2}
Ghosh, P., \& Lamb, F.K. 1979, ApJ {\bf 232}, 256
\bibitem{3}
Konenkov, D.Yu., \& Popov, S.B. 1997, PAZh, {\bf 23}, 569
\bibitem{4}
Lipunov, V.M., 1992, ``Astrophysics of Neutron Stars'',
Springer--Verlag (Berlin)
\bibitem{5}
Lipunov, V.M., \& Popov, S.B. 1995a, AZh, {\bf 72}, 711
\bibitem{6}
Lipunov, V.M., \&  Popov, S.B. 1995b,
  Astron. Astroph. Transactions,  {\bf 8}, 221  
\bibitem{7}Motch, C. 2000, astro-ph/0008485
\bibitem{8}Prokhorov, M.E., Popov, S.B., \& Khoperskov, A.V. 2001,
astro-ph/0108503
\bibitem{9}
Treves, A., Turolla, R., Zane, S., \& Colpi, M. 2000,
PASP {\bf 112}, 297
\end{thebibliography}
\end{document}